\documentclass[preprint,showpacs,preprintnumbers,amsmath,amssymb,nofootinbib]{revtex4}
\usepackage{graphicx}
\usepackage{subfigure}
\usepackage{dcolumn}
\usepackage{bm}

\def \be {\begin{equation}}
\def \ee {\end{equation}}
\def \bea{\begin{eqnarray}}
\def \eea{\end{eqnarray}}

\def \prd{{\it Phys. Rev. D, }}

\def \apj{{\it Ap. J., }}

\def \CQG{{\it Class. Quantum Grav., }}

\begin{document}
\title{Single-detector searches for a stochastic background of
  gravitational radiation}

\author{Massimo Tinto} 
\email{Massimo.Tinto@das.inpe.br}
\affiliation{Instituto Nacional de Pesquisas Espaciais, Astrophysics
  Division, Avenida dos Astronautas 1758, 12227-010 - S\~{a}o Jos\'{e} dos
  Campos, SP, Brazil}
\altaffiliation [Also at: ]{Jet Propulsion Laboratory, California
  Institute of Technology, Pasadena, CA 91109}
\author{J.W. Armstrong}
\email{John.W.Armstrong@jpl.nasa.gov}
\affiliation{Jet Propulsion Laboratory, California Institute of
  Technology, Pasadena, CA 91109}

\date{\today}

\begin{abstract}
  We propose a data processing technique that allows searches for a
  stochastic background of gravitational radiation with data from a
  single detector. Our technique exploits the difference between the
  coherence time of the gravitational wave (GW) signal and that of the
  instrumental noise affecting the measurements. By estimating the
  auto-correlation function of the data at an off-set time that is
  longer than the coherence time of the noise {\underbar {but}}
  shorter than the coherence time of the GW signal, we can effectively
  enhance the power signal-to-noise ratio (SNR) by the square-root of
  the integration time.  The resulting SNR is comparable in magnitude
  to that achievable by cross-correlating the data from two co-located
  and co-aligned detectors whose noises are uncorrelated. Our method
  is general and it can be applied to data from ground- and
  space-based detectors, as well as from pulsar timing experiments.
\end{abstract}

\maketitle

\section{Introduction}
\label{intro}

The direct detection of gravitational radiation is one of the most
pressing challenges in the physics of this century.  Predicted by
Einstein shortly after formulating his general theory of relativity,
gravitational waves (GW) will allow us to probe regions of space-time
otherwise unobservable in the electromagnetic spectrum
\cite{Thorne1987}.  Several experimental efforts have been underway
for awhile, both on the ground and in space
\cite{LIGO,VIRGO,GEO,TAMA,DOPPLER,PPA98,Pulsar,JAT2011}, and only
recently kilometer-size ground-based interferometers have been able to
identify the most stringent upper-limits to date for the amplitudes of
the radiation expected from several classes of GW sources.  Although
present-generation instruments have not been able yet to unambiguously
observe GWs, next-generation Earth-based interferometers and
pulsar-timing experiments \cite{Pulsar,JAT2011}, as well as newly
envisioned space-based detectors \cite{PPA98}, are expected to achieve
this goal.

The physical characteristics of the astrophysical sources emitting
gravitational radiation varies significantly over the observational
frequency bands of these detectors. Depending on the specific detector
design, we will observe GWs emitted during cataclysmic events such as
supernova explosions and coalescences of binary systems containing
neutron stars and black-holes, spiraling white-dwarfs binaries in our
Galaxy and binary black-holes across the Universe, and a stochastic
background of astrophysical or cosmological origin \cite{Thorne1987}.

The detection of a stochastic background of cosmological origin, in
particular, will allow us to infer properties about the very early
Universe and probe various cosmological models predicting the
formation mechanism of the background.  Since the characteristic
strength of such stochastic backgrounds is expected to be smaller than
the instrumental noise level of present and of foreseeable future
detectors, it is imperative to identify data processing techniques
capable of enhancing the likelihood of detection. 

The most general and robust of such techniques relies on
cross-correlating the data from pairs of detectors operating in
coincidence and whose noises are uncorrelated. By virtue of observing
the same GW signal while being affected by noise sources that are
uncorrelated, the cross-correlation technique increases the resulting
SNR to a stochastic background by the square-root power of the
integration time. By relying on a network of $N$ operating detectors,
the overall sensitivity can be further enhanced by a factor that
depends on $N$ \cite{AllenRomano1999}.

Although the cross-correlation technique for detecting a
background of gravitational radiation can be applied to ground-based
detectors as they are large in number and widely separated on Earth
(so their noises can be regarded as uncorrelated), it is presently
unthinkable to extend it to space-based interferometers as only one
might become operational in the next decade. Although it might be
argued that space-based interferometers such as LISA, for example, can
generate up to three Time-Delay Interferometric (TDI) combinations
\cite{TD2005} that could be cross-correlated, it is easy to see that
their noises would be correlated making the cross-correlation
technique ineffective (in simple terms, a TDI combination has some of
its noises correlated to those affecting the other TDI responses by
virtue of sharing with them at least ``one-arm'').  \footnote{Although
  it is possible to construct, within the TDI space, three TDI
  combinations whose noises are uncorrelated \cite{PTLA2002}, it is
  straightforward to show that also their responses to an isotropic
  stochastic background of gravitational radiation would be
  uncorrelated.}

These considerations imply that a data processing technique for
detecting a background of gravitational radiation with a
single-detector response is required. Although some work in this
direction has already appeared in the literature for the LISA mission
and based on the use of the ``null-stream'' data combination
\cite{TAE2000,RRV2008}, further work is still needed. It is within
this perspective that the present article proposes a new approach to
this problem based on the use of the auto-correlation function.  Our
method provides a statistic for enhancing the likelihood of detection
of a GW stochastic background with a single detector.  As we will show
below, the resulting SNR is comparable to that obtainable by
cross-correlating the data from two co-located and co-aligned
detectors whose noises are assumed uncorrelated.

The paper is organized as follows.  In section \ref{Auto} we derive
the expression of the SNR associated with the auto-correlation
function of the data, estimated at an off-set time $\tau$ that is
longer than the coherence time of the instrumental noise but shorter
than that of the stochastic background. After noticing that the SNR
can be further enhanced by introducing a weighting function,
$Q(\tau)$, in the auto-correlation function, we derive the expression
for $Q$ that results into an optimal SNR. Finally in section
\ref{Conclusions} we present our comments and conclusions, and
emphasize that our auto-correlation technique can be applied not only
to the data from a forthcoming space-based interferometers, but also
to those from a network of ground-based interferometers and pulsar
timing.

\section{The auto-correlation function}
\label{Auto}

Let us denote with $R(t)$ the time-series of the detector output data,
which we will assume to include a random process, $n (t)$, associated
with the noise and also a random process associated with a GW
stochastic background, $h(t)$, in the following form
\begin{equation}
R(t) \equiv h(t) + n(t) \ .
\label{response}
\end{equation}
In what follows we will assume the GW stochastic background to be an
unpolarized, stationary, and Gaussian random process with
zero mean. As a consequence of these assumptions such a background is
characterized by a one-sided power spectral density, $P_h
(|f|)$  defined by the following expression \cite{AllenRomano1999,Papoulis2002}
\begin{equation}
\langle {\widetilde {h}} (f) \ {\widetilde {h^*}} (f') \rangle \equiv
\frac{1}{2} \delta (f - f') \ P_h (|f|) \ ,
\label{HSpectra}
\end{equation}
where the symbol $ \ {\widetilde {} } \ $ represents the operation of
Fourier transform, the $^*$ symbol denotes the usual operation of
complex conjugation, and the angle-brackets, $\langle \rangle$, denote
the operation of ensemble average of the random
process.  We will further assume $n$ to be stationary, Gaussian
distributed with zero-mean, and characterized by a one-sided power
spectral density, $P_n(|f|)$, defined as follows
\begin{equation}
\langle {\widetilde {n}} (f) \ {\widetilde {n^*}} (f') \rangle \equiv
\frac{1}{2} \delta (f - f') \ P_n (|f|) \ .
\label{NSpectra}
\end{equation}

Let us denote with $\tau_{\rm gw}$ and $\tau_n$ the coherence time of
the GW stochastic background and of the instrumental noise
respectively, and let us also assume $\tau_{\rm gw} > \tau_n$.  This
assumption is rather general since, by being true for white noise and
any colored GW background, it must remain true in general as the data
can always be pre-whitened \footnote{Although the case when both the
  noise and the stochastic background spectra are white over the
  entire observational band cannot be tackled by our technique,
  theoretical models for likely backgrounds of cosmological and
  astrophysical origins characterize them with ``colored'' spectra
  that are in general different from those associated with the noises
  of the detector \cite{Thorne1987}}.

\begin{figure}
\centering
\includegraphics[width=6.5 in, angle = 0]{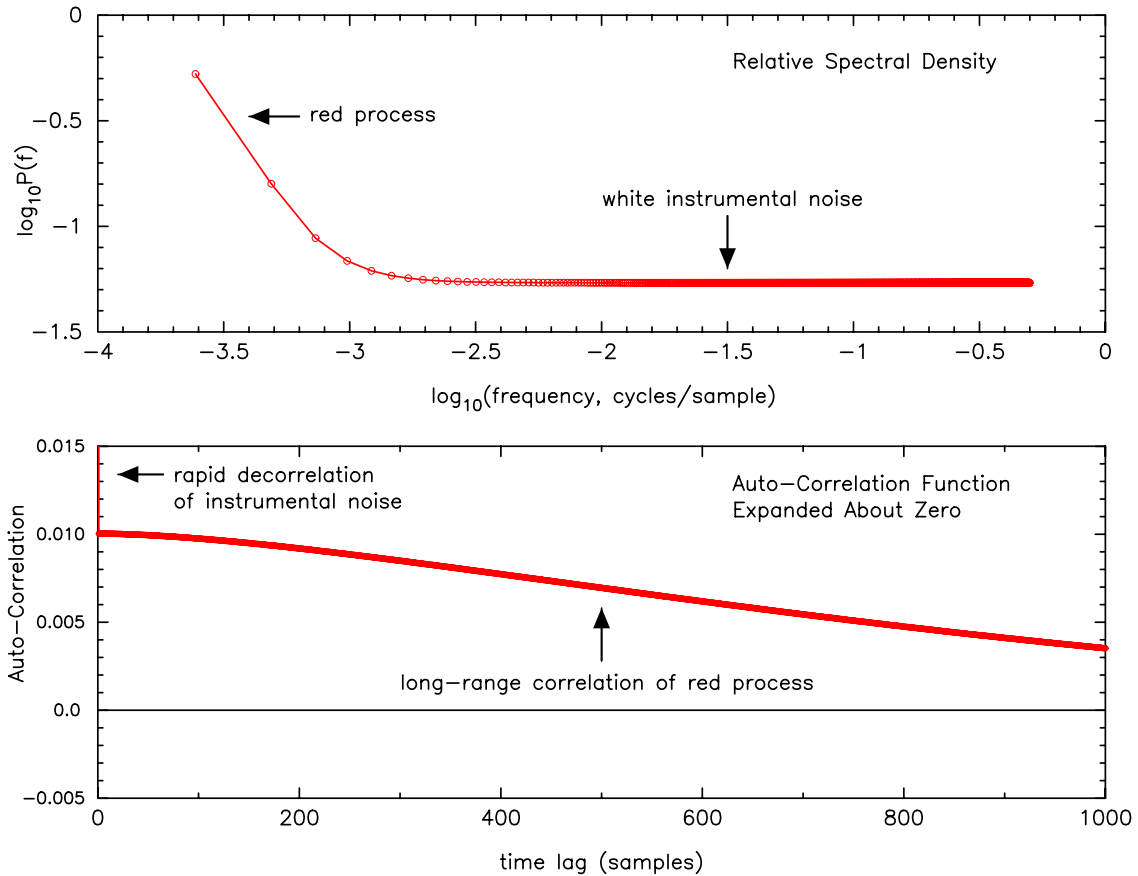}
\caption{Schematic illustration of the method.  The upper panel shows
  the power spectrum of (assumed-white) instrumental noise plus a
  low-frequency excess ("red") signal power. The lower panel shows the
  auto-correlation function of the spectrum in the upper panel,
  expanded around zero correlation.  The white component decorrelates
  rapidly from unity, while the red process gives rise to long-range
  correlations.  The detection of these correlations competes with
  estimation error statistics of the correlation function, with an SNR
  increasing like $T^{1/2}$, as described in the text.}
\label{figure:cartoon}
\end{figure}

The sample auto-correlation function, $S (\tau)$, of the data, $R(t)$,
can be written in the following form
\begin{equation}
S (\tau) \equiv \int^{T/2}_{-T/2} R(t) \ R(t + \tau) \ dt \ ,
\label{correla}
\end{equation}
where $T$ is the time interval in seconds over which the
auto-correlation is computed. The interval of interest for our method
is when $\tau_n < \tau < \tau_{\rm gw}$; in this region the
instrumental noise has decorrelated, leaving the correlation function
of the signal.  Figure~\ref{figure:cartoon} shows the situation
schematically.  This signal correlation then competes with the
estimation error statistics of the correlation function for detection,
as discussed below.  Very similarly to the cross-correlation technique
discussed in \cite{AllenRomano1999,Mitraetal2008} we can rely on $S
(\tau)$ for building a decision rule for the detection of a GW
stochastic background.

From the statistical properties of both the GW stochastic background
and the noise, and by taking into account the central-limit theorem
(for $\tau_{\rm gw} << T$, $S(\tau)$ is equal to the sum of many
uncorrelated products) it follows that $S(\tau)$ is also a Gaussian random
process \cite{Papoulis2002}. In order to fully characterize it we can
estimate its mean and variance, which must be functions of only the
spectra of the two random processes $h(t)$ and $n(t)$.

By taking the ensemble average of $S$ and $S^2$ over many noise and GW
signal realizations and for $T >> \tau_{\rm gw}$, after some long but
straightforward calculations we get the following expressions for the
mean and variance of $S (\tau)$ (see Appendix \ref{appendixA} for
details)\footnote{It should be noticed that Eq. (\ref{Variance}) can
  also be derived from the equation for the estimation error of the
  auto-correlation function given in Jenkins and Watts \cite{JW68}
  (eq. 5.3.20), under the approximation that $T >> \tau_{\rm gw}$, and
  that the width of the square of the auto-correlation function is
  small compared with $T$.}
\begin{equation}
\mu_S \equiv \langle S \rangle = \frac{T}{2} \ \int^{\infty}_{-\infty} P_h (|f|)
e^{2 \pi i f \tau} \ df \ ,
\label{Mean}
\end{equation}
\begin{equation}
\sigma^2_S \equiv \langle S^2 \rangle - \langle S \rangle^2
= \frac{T}{2} \ 
\int^{\infty}_{-\infty} \cos^2(2 \pi f \tau) \ [P_h (|f|) + P_n
(|f|)]^2 \ df \ .
\label{Variance}
\end{equation}
Eqs.(\ref{Mean}, \ref{Variance}) can then be used for deriving the
following expression for the Signal-to-Noise Ratio ($SNR_{auto}$)
\begin{equation}
SNR_{auto} \equiv \frac{\mu_S}{\sigma_S} =  
\sqrt{\frac{T}{2}} \ \frac{\int^{\infty}_{-\infty} P_h (|f|)
e^{2 \pi i f \tau} \ df}{\left[\int^{\infty}_{-\infty} \cos^2(2 \pi f \tau) \ [P_h (|f|) + P_n
(|f|)]^2 \ df \right]^{1/2}} \ .
\label{SNRauto}
\end{equation}

In order to get a quantitative understanding of the above expression
for the SNR of the auto-correlation at lag $\tau$, we may compare it
against the SNR associated with the cross-correlation (computed also
at time-lag $\tau$) of the data from two co-aligned and co-located
detectors whose noises are assumed to be uncorrelated (see
\cite{AllenRomano1999} for details). It is straightforward to obtain
the following expression for the cross-correlation signal-to-noise
ratio $SNR_{2d}$
\begin{equation}
SNR_{2d} = \sqrt{\frac{T}{2}} \ \frac{\int^{\infty}_{-\infty} P_h (|f|)
e^{2 \pi i f \tau} \ df}{\left[\int^{\infty}_{-\infty} 
\left[P_h^2 (|f|) \cos^2(2 \pi f \tau) + \frac{1}{2} P_h (|f|) \ (P_{1n}
(|f|) + P_{2n}(|f|)) + \frac{1}{2} P_{1n} (|f|) 
P_{2n}(|f|)\right]\right]^{1/2}} \ ,
\label{SNR2d}
\end{equation}
where we have denoted with $P_{in}(|f|)$ the one-sided power spectral
density of the noise in detector $i = 1, 2$. In order to compare the
two expressions given in Eqs. (\ref{SNRauto}, \ref{SNR2d}), let us
assume the two noise power spectral densities $P_{in}$ to be equal to
each other and to a white spectrum having spectral level equal to $A$
per unit bandwidth. By taking the ratio between $SNR_{2d}$ and
$SNR_{auto}$ we get
\begin{equation}
\frac{SNR_{2d}}{SNR_{auto}} = \frac{\left[\int^{\infty}_{-\infty} 
\cos^2(2 \pi f \tau) \ [P_h (|f|) + A]^2 \ df \right]^{1/2}}
{\left[\int^{\infty}_{-\infty} 
\left[P_h^2 (|f|) \cos^2(2 \pi f \tau) + P_h (|f|) \ A + \frac{1}{2} A^2
\right]\right]^{1/2}} \ .
\label{ratio}
\end{equation}
Since current estimates for the likely amplitudes of cosmological
gravitational wave backgrounds indicate them to be significantly
smaller than the instrumental noise levels of current and future
detectors, if we take $P_h << A$ in Eq. (\ref{ratio}) 
in this limit we find $SNR_{2d}/ SNR_{auto} \rightarrow
1$.\footnote{It is easy to show that, also in the limit of ``high
  signal-to-noise ratio'', $P_h >> A$, $SNR_{2d}/ SNR_{auto}
  \rightarrow 1$.}

\subsection{The optimal filter}
\label{Optimal}

Although the physical reasons underlying the selection of the time
interval within which the time-lag $\tau$ should be selected are quite
clear, they do not allow us to uniquely identify a value for it.  In
particular, since the number of possible time lags we could use for
estimating the auto-correlation function can be in general rather
large, it is natural to look for the optimal way of combining all
possible auto-correlation functions. In analogy with the
cross-correlation technique \cite{AllenRomano1999} in which an optimal
filter $Q(t, t')$ maximizing the SNR could be identified, here we show
that also for the auto-correlation technique there exists such a
filter. Since the stochastic background and the instrumental noise
have been assumed to be stationary, we introduce a time-invariant
filter function and write the following linear combination of all
possible sample auto-correlation functions lags
\begin{equation}
{\mathcal{S}} \equiv \int^{T/2}_{-T/2} \ \int^{+\infty}_{-\infty} \ R(t) R(t +
\tau) Q(\tau) \ dt \ d\tau \ ,
\label{SS}
\end{equation}
where the auto-correlation is computed over the time interval $T$, and
the filter function $Q(\tau)$ at the time-lag $\tau$ is non-zero in
the interval $\tau_n < |\tau| < \tau_{gw}$ and zero everywhere else (a
property that we have already taken into account in Eq. (\ref{SS}) by
extending the $\tau$-integration over the entire real axis).

From the statistical properties of both the GW stochastic background
and the noise, and by virtue of the central-limit theorem, it follows
that $\mathcal S$ is also a Gaussian random process \cite{Papoulis2002} that
can be fully characterized by estimating its mean $\mu_{\mathcal S}$ and
variance, $\sigma^2_{\mathcal S}$. Although their derivations are quite long, they
are straightforward and result into the two following expressions
\begin{equation}
\mu_{\mathcal S} = \frac{T}{2} \int^{+\infty}_{-\infty} P_h (|f|) \ {\widetilde
  Q} (f) \ df \ ,
\label{mus}
\end{equation}
\begin{equation}
\sigma^2_{\mathcal S} = \frac{T}{2} \int^{+\infty}_{-\infty} [P_h (|f|) + P_n
  (|f|)]^2 \ |{\widetilde Q} (f)|^2 \ df \ .
\label{sigmas}
\end{equation}

By simple inspection of Eq. (\ref{sigmas}), and following
\cite{AllenRomano1999}, it is convenient to define the following
operation of inner product between two arbitrary complex functions,
say $A$, and $B$
\begin{equation}
(A,B) \equiv \int^{+\infty}_{-\infty} A(f) B^* (f) [P_h (|f|) + P_n
(|f|)]^2 \ df \ .
\label{inner}
\end{equation}
With this newly defined inner product, $\mu_{\mathcal S}$ and $\sigma_{\mathcal S}$ can be
rewritten in the following form
\begin{equation}
\mu_{\mathcal S} = \frac{T}{2} ({\widetilde Q}, \frac{P_h}{[P_h + P_n]^2}) \ \ \
\ ; \ \ \ \ \sigma^2_{\mathcal S} = \frac{T}{2} (Q, Q) \ ,
\end{equation}
while the squared signal-to-noise ratio of $\mathcal S$, $SNR^2$ is equal to 
\begin{equation}
SNR^2 \equiv \frac{\mu_{\mathcal S}^2}{\sigma^2_{\mathcal S}} = \frac{T}{2} \ 
\frac{ ({\widetilde Q}, \frac{P_h}{[P_h + P_n]^2})^2 }{(Q, Q)} \ .
\label{SNRP}
\end{equation}
From a simple geometrical interpretation of the inner product defined
through Eq. (\ref{inner}), from Eq. (\ref{SNRP}) it is easy to see
that the maximum of the SNR of $\mathcal S$ is achieved by choosing the filter
function ${\widetilde Q}$ to be equal to
\begin{equation}
{\widetilde Q} (f) = \lambda \ \frac{P_h (|f|)}{[P_h(|f|) +
  P_n(|f|)]^2} \ ,
\label{Q}
\end{equation}
where $\lambda$ can be any arbitrary real number.  The above
expression for the filter $Q$ implies the following maximum SNR
achievable by the auto-correlation technique
\begin{equation}
SNR^2_{max} = \frac{T}{2} \int^{+\infty}_{-\infty}
\frac{P_h^2(|f|)}{[P_h (|f|) + P_n (|f|)]^2} \ df \ .
\end{equation}

\section{Discussions and conclusions}
\label{Conclusions}

The main result of our work has been to show that the auto-correlation
function of the data measured by a single detector of gravitational
radiation can be used for searching for a stochastic gravitational
wave background. This is possible when there is a difference between
the spectral properties of the GW background and of the instrumental
noise. By estimating the auto-correlation function of the data at a
time-lag at which the auto-correlation of the noise has decayed to a
value smaller than that of the auto-correlation function of the GW
background, one can enhance the resulting power signal-to-noise ratio
by the square-root of the integration time.

We have assessed the effectiveness of our technique by comparing the
expression of its SNR against that achievable by cross-correlating the
data from two co-located and co-aligned detectors whose noises are
assumed to be uncorrelated. We found the two expressions to coincide
in the expected limit of the noise power spectrum larger than
that of the GW background.  In order to further enhance the SNR
associated with the auto-correlation technique we have derived the
expression of an optimal filter function that reflects any prior
knowledge we might have about the spectrum of GW background.

The auto-correlation technique can be used when searching for a
stochastic background of gravitational radiation with (i) networks of
ground-based interferometer as it will further enhance the likelihood
of detection achievable with the cross-correlation method, (ii) with a
single space-based interferometer, and (iii) with pulsar timing data.

\section*{Acknowledgments}

M.T. acknowledges FAPESP for financial support (Grant
No. 2011/11719-0). For J.W.A. this research was performed at the Jet
Propulsion Laboratory, California Institute of Technology, under
contract with the National Aeronautics and Space Administration.(c)
2012 California Institute of Technology.  Government sponsorship
acknowledged.

\appendix
\section{Derivation of the mean and variance of $S(\tau)$}
\label{appendixA}

In what follows we derive the expressions of the mean and variance of
the auto-correlation function $S(\tau)$ (Eqs. (\ref{Mean},
\ref{Variance}) given in Section (\ref{Auto})). We will assume the
instrumental noise to be characterized by a random process $n(t)$ that
is Gaussian-distributed with zero-mean, and has one-sided
power-spectral density $P_n(|f|)$. We will also take the GW stochastic
background to be a Gaussian random process of zero-mean, unpolarized,
and stationary.  As a consequence of these assumptions such a
background is uniquely characterized by a one-sided power spectral
density, $P_h (|f|)$ as given by Eq.  (\ref{HSpectra}).

From the definitions of the data, $R(t)$, and of the auto-correlation
function, $S (\tau)$, calculated at a time-lag $\tau$ such that
$\tau_n < \tau < \tau_{\rm gw}$, we have
\begin{equation}
S (\tau) \equiv \int^{T/2}_{-T/2} [h(t) + n(t)] \ [h(t + \tau) + n(t +
\tau)]\ dt \ ,
\label{correlaA}
\end{equation}
where $T$ is the time interval in seconds over which the
auto-correlation function is computed and it is assumed to be longer
than $\tau_{\rm gw}$. After taking the ensemble average of both sides
of Eq. (\ref{correlaA}), we get
\begin{equation}
\langle S (\tau) \rangle = \int^{T/2}_{-T/2} \left[\langle h(t) h(t + \tau) \rangle 
+ \langle h(t) n(t + \tau) \rangle + \langle n(t) h(t + \tau) \rangle 
+ \langle n(t) n(t + \tau) \rangle \right] \ dt \ .
\label{correlaB}
\end{equation}
Note that the second and third terms on the right-hand-side of
Eq. (\ref{correlaB}) are equal to zero because the random process
associated with the stochastic GW background and that associated with
the instrumental noise are uncorrelated. The fourth term instead can
also be regarded to be zero because is the auto-correlation
function of the noise estimated at an off-set time longer than its
coherence time. From these considerations we find the following
expression for the mean of $S$
\begin{equation}
\mu_S \equiv \langle S (\tau) \rangle = \int^{T/2}_{-T/2} \left[\langle h(t) h(t + \tau) \rangle 
\right] \ dt \ = \frac{T}{2} \int_{-\infty}^{+\infty} P_h(|f|)
e^{2\pi i f \tau} \ df \ .
\label{correlaC}
\end{equation}
In order to derive the expression for the variance of $S$
(Eq. (\ref{Variance})) we need to calculate $\langle S^2
\rangle$. Since $S^2 (\tau)$ can be written as
\begin{eqnarray}
S^2 (\tau) & = & \int^{T/2}_{-T/2} \int^{T/2}_{-T/2} 
\left[h(t) h(t + \tau) + h(t) n(t + \tau) + n(t) h(t + \tau) + 
n(t) n(t + \tau) \right]
\nonumber
\\
& \times &
\left[h(t') h(t' + \tau) + h(t') n(t' + \tau) + n(t') h(t' + \tau) + 
n(t') n(t' + \tau) \right] \ dt \ dt' \ ,
\label{S2}
\end{eqnarray}
by taking the ensemble average of both sides of Eq. (\ref{S2}) only
nine of the sixteen terms present are different from zero. This
follows from the Gaussian nature of both random processes and because
terms such as $\langle x(t) y(t') z(t + \tau) w(t' + \tau) \rangle$
(where $x, y, z, w$ can be either $h$ or $n$) can be expressed in the
following form \cite{Papoulis2002}
\begin{eqnarray}
\langle x(t) y(t') z(t +
\tau) w(t' + \tau) \rangle & = & \langle x(t) y(t') \rangle \langle z(t +
\tau) w(t' + \tau) \rangle + \langle x(t) z(t + \tau) \rangle \langle
y(t') w(t' + \tau) \rangle 
\nonumber
\\
& + & \langle x(t) w(t' + \tau) \rangle \langle
y(t') z(t + \tau) \rangle 
\end{eqnarray}
From the above considerations we get the following expression for $\langle S^2 (\tau) \rangle$
\begin{eqnarray}
\langle S^2 (\tau) \rangle & = & \int^{T/2}_{-T/2} \int^{T/2}_{-T/2} 
\left[
\langle h(t) h(t + \tau) \rangle \langle h(t') h(t' + \tau) \rangle +
\langle h(t) h(t') \rangle \langle h(t + \tau) h(t' + \tau) \rangle \right. 
\nonumber
\\
& + & \left.
\langle h(t) h(t' + \tau) \rangle \langle h(t') h(t + \tau) \rangle +
\langle h(t) h(t') \rangle \langle n(t + \tau) n(t' + \tau) \rangle \right.
\nonumber
\\
& + & \left.
\langle h(t) h(t' + \tau) \rangle \langle n(t + \tau) n(t') \rangle +
\langle h(t') h(t + \tau) \rangle \langle n(t) n(t' + \tau) \rangle
\right. 
\nonumber
\\
& + & \left.
\langle h(t + \tau) h(t' + \tau) \rangle \langle n(t) n(t') \rangle +
\langle n(t) n(t') \rangle \langle n(t + \tau) n(t' + \tau) \rangle
\right.
\nonumber
\\
& + & \left.
\langle n(t) n(t' + \tau) \rangle \langle n(t') n(t + \tau) \rangle
\right]  \ dt \ dt' \ , 
\label{S21}
\end{eqnarray}
which can further be rewritten in the following form
\begin{eqnarray}
\langle S^2 (\tau) \rangle & = & 
\frac{T^2}{4} \int^{+\infty}_{-\infty} \int^{+\infty}_{-\infty} 
P_h(|f|)P_h(|f'|) \ e^{2\pi i (f - f')\tau} \ df \ df'
+
\frac{T}{4} \int^{+\infty}_{-\infty} P^2_h(|f|) \ df  
\nonumber
\\
& + & 
\frac{T}{4} \int^{+\infty}_{-\infty} P^2_h(|f|) e^{4\pi i f\tau} \ df 
+ \frac{T}{4} \int^{+\infty}_{-\infty} P_h(|f|) P_n(|f|) \ df 
\nonumber
\\
& + & 
\frac{T}{4} \int^{+\infty}_{-\infty} P_h(|f|) P_n(|f|) e^{4 \pi i f
  \tau} \ df 
+
\frac{T}{4} \int^{+\infty}_{-\infty} P_h(|f|) P_n(|f|) e^{-4 \pi i f
  \tau} \ df 
\nonumber
\\
& + & 
\frac{T}{4} \int^{+\infty}_{-\infty} P_h(|f|) P_n(|f|) \ df 
+ \frac{T}{4} \int^{+\infty}_{-\infty} P^2_n(|f|) \ df 
\nonumber
\\
& + & 
\frac{T}{4} \int^{+\infty}_{-\infty} P^2_n(|f|) e^{4 \pi i f \tau} \
df \ .
\label{S22}
\end{eqnarray}
If we now square both sides of Eq. (\ref{correlaC}) and subtract it
from Eq. (\ref{S22}) to obtain the expression for the variance
$\sigma^2_S$, after some rearrangement of the remaining terms we get
the following result
\begin{equation}
\sigma^2_S = \frac{T}{2} \int^{+\infty}_{-\infty} \cos^2(2 \pi f \tau)
[P_h(|f|) + P_n(|f|)]^2 \ df \ .
\end{equation}

\end{document}